\documentclass[journal abbreviation, manuscript]{copernicus}

\usepackage{amsthm}
\usepackage{natbib}
\usepackage{physics}
\usepackage{float}

\begin{document}
\nolinenumbers
\title{PyDDC: An Eulerian-Lagrangian simulator for density driven
convection of $\mathrm{CO_2}$–-brine systems in saturated porous media}

% \Author[affil]{given_name}{surname}
\Author[1]{Sayan}{Sen}
\Author[1][skh@bgu.ac.il]{Scott K.}{Hansen} %% correspondence author

\affil[1]{Zuckerberg Institute for Water Research, Ben-Gurion University of the Negev, Midreshet
Ben-Gurion, Israel}
% \affil[1]{ADDRESS}

%% The [] brackets identify the author with the corresponding affiliation. 1, 2, 3, etc. should be inserted.

%% If an author is deceased, please mark the respective author name(s) with a dagger, e.g. "\Author[2,$\dag$]{Anton}{Smith}", and add a further "\affil[$\dag$]{deceased, 1 July 2019}".

%% If authors contributed equally, please mark the respective author names with an asterisk, e.g. "\Author[2,*]{Anton}{Smith}" and "\Author[3,*]{Bradley}{Miller}" and add a further affiliation: "\affil[*]{These authors contributed equally to this work.}".

\runningtitle{TEXT}

\runningauthor{TEXT}

\received{}
\pubdiscuss{} %% only important for two-stage journals
\revised{}
\accepted{}
\published{}

%% These dates will be inserted by Copernicus Publications during the typesetting process.

\firstpage{1}

\maketitle

\begin{abstract}
PyDDC is a particle tracking reservoir simulator capable of solving non-linear density driven convection of single phase carbon-dioxide ($\mathrm{CO_2}$)--brine fluid mixture in saturated porous media at the continuum scale. In contrast to the sate-of-the-art Eulerian models, PyDDC uses a Lagrangian approach to simulate the Fickian transport of single phase solute mixtures. This introduces additional flexibility of incorporating anisotropic dispersion and benefits from having no numerical artifacts in its implementation. It also includes $\mathrm{CO_2}$--brine phase equilibrium models, developed by other researchers, to study the overall dynamics in the presence of electrolyte brine at different pressure and temperatures above the critical point of $\mathrm{CO_2}$. We demonstrate the implementation procedure in depth, outlining the overall structure of the numerical solver and its different attributes that can be used for solving specific tasks.
\end{abstract}

% \copyrightstatement{TEXT} %% This section is optional and can be used for copyright transfers.
\introduction  %% \introduction[modified heading if necessary]
Simulating fluid flow and transport in saturated porous media has gained considerable attention over the past few decades owing to its different industrial applications, predominantly in the field of geological carbon storage \citep{riaz2014carbon, meng2014numerical, chen2023density},  subsurface energy transport \citep{provost2019sutra, karvounis2011modeling, soboleva2018density} and salt-water intrusion \citep{seng2021impacts, kalakan2018saltwater, smith2004mixed}. Qualitative analyses of geological storage process is challenging as the overall flow dynamics is highly non-linear and requires sophisticated numerical methods to be invoked that are capable of capturing the complexities associated with most flows in natural aquifers. In the context of $\mathrm{CO_2}$-sequestration, the important quantitative aspects analyzed are the onset of convection, mixing or dissolution rate and regime transitions marked by evolution of high density fingering structures. Practically all numerical solvers developed in this regard are fully Eulerian. Both solute transport software like GEOSX \citep{osti_1422506}, DuMux3 \citep{koch2021dumux}, PFLOTRAN \citep{lichtner2015pflotran}, MT3D-USGS \citep{bedekar2016mt3d}, Hydrus-2D/3D \citep{varvaris2021parameterization} and high-performance open-source general purpose multi-physics software like FEniCS \citep{logg2012automated}, COMSOL \citep{liu2023numerical}, OpenFOAM \citep{chu2016investigation} have been used for solving fluid mechanics problems in porous media with GEOSX being dedicated primarily to $\mathrm{CO_2}$ storage projects. Other works include direct numerical simulation (DNS) approaches employing full scale Eulerian solvers and perturbation based Linear Stability Analysis models \citep{hansen2023impacts, soltanian2016critical, tsinober2022role, michel2020role, kong2013numerical, hidalgo2009effect, farajzadeh2010effect, de2021influence, hewitt2013convective, emami2015convective, green2018steady, hassanzadeh2007scaling, riaz2006onset, emami2017stability} to understand the processes responsible for generating instability and characterize the flow dynamics of $\mathrm{CO_2}$--brine mixtures in porous media at different reservoir heterogeneities. These studies, along with many others, have been pivotal in developing our knowledge about $\mathrm{CO_2}$ plume behaviour in the subsurface. 
Although Eulerian flow solvers have been used extensively, they have inherent problems with handling flows having high advective velocities and sharp fronts. A naive implementation in such case results in strong dispersive effects giving rise to un-physical oscillations in the target field variable. For handling numerical dispersion higher order Eulerian schemes like TVD (Total Variation Diminishing) \citep{darwish2003tvd} and WENO (Weighted Essentially Non-Oscillatory) schemes \citep{capdeville2008central} have been developed, but those methods are computationally-demanding when it comes to solving large scale flow and transport problems. 

Lagrangian approaches offer greater flexibility when it comes to simulating transport phenomena. \cite{van2018lagrangian} presents a thorough review about the different types of particle based methods that are used for capturing sub-scale diffusion and unresolved physics. These techniques, in general, have particularly proved it worth for modelling heterogeneous flow systems pertaining to both Fickian and non-Fickian physics \citep{kitanidis1994particle, fernandez2005differences, delay2005simulating, srinivasan2010random, hansen2018direct}. Inherently, particle based methods are by nature free of numerical artifacts like spurious dispersion and are more efficient than Eulerian methods for capturing sharp concentration gradients and sub-scale transport behaviour. Particle trackers have seen use for subsurface transport in linear systems where the flow regime does not depend on transport properties. Aurora \citep{hansen2020aurora} and EcoSLIM \citep{maxwell2019exploring} are two examples of particle tracking software that are used in solute transport applications for linear flow problems. In the context of $\mathrm{CO_2}$ sequestration, which features density-driven flow and thus bi-directional coupling between flow and transport, there have been no such particle based techniques reported that has been used to simulate $\mathrm{CO_2}$ transport behaviour in the subsurface. Problems involving bi-directional coupling requires the flow field and the scalar concentration to be computed at each simulation step giving rise to a combined Euler-Lagrange scheme, in contrast to the straightforward Lagrangian trackers that involve only one-way coupling.

In Euler-Lagrange (EL) method, where there is a feedback between flow and transport solver, a particle to grid projection operation and vice versa is required to establish the two-way coupling. Kernel based methods used for these projection operations allow one to carry out different levels of spatial interpolation by using different types of interpolating kernels \citep{bagtzoglou1992projection}. The transport part in EL methods can either be solved using tracer based model, where the advective transport is solved using particle based method while the diffusive or dispersive transport is solved on a discretized grid using the Eulerian approach, or using the stochastic model, where advective and dispersive terms are not decoupled but are included together in a single equation called the \textit{stochastic differential equation} (SDE) and is solved using the method of random walk particle tracking (RWPT). \cite{neuman1982eulerian} also discussed different EL schemes explaining their applicability for advection-dominated and dispersion-dominated scenarios. \cite{maljaars2021leopart} recently developed a tracer based Euler-Lagrange method to solve general-purpose Rayleigh-Taylor instability problems enforcing mass conservation by using a PDE-constrained optimization procedure. However, their approach is not readily adapted for $\mathrm{CO_2}$--brine systems. For simulating coastal dispersion, \cite{suh2006hybrid} combined both EL and RWPT approaches to reduce the computational complexity arising from the implementation of only the random walk methods. Stochastic model is the preferred choice in our case because of the ease of RWPT scheme to enforce conservation properties and handle steep concentration gradients compared to the tracer based model. The particle tracking based transport solver that we develop is novel in this regard as it integrates a Eulerian flow solver with a Lagrangian based transport module in a computationally efficient manner that can be used to solve density-driven convection and understand instabilities originating from $\mathrm{CO_2}$--brine miscible displacement.

\section{Theory}

Supercritical $\mathrm{CO_2}$ (Sc-$\mathrm{CO_2}$) upon injection into the deep saline aquifer is less dense than the ambient brine and due to the effect of buoyancy it rises and accumulates at a stratigraphically higher level when it encounters a geological barrier. Over time the $\mathrm{CO_2}$ diffuses into the underlying brine and gives rise to a $\mathrm{CO_2}$--brine diffusive layer that has a characteristic density larger than the density of the underlying brine. This results in an unstable configuration having a density stratification with a denser medium overlying a less dense one. Subsequently variations in the flow field introduced by the medium heterogeneity gives rise to small scale perturbations in the diffusive layer and as a response to those perturbations the diffusive layer breaks down in the form of dense fingers which propagates downwards at advective rates. 

For density-driven convection in porous media, the coupled non-linear flow and transport equations for single-phase miscible fluids are given by:

\begin{gather}
\label{eq1}
    \Vec{q} = -\frac{k(x)}{\mu(c)}(\nabla P + \rho(c) g \Vec{e}_z) 
    \\
    \label{eq2}
    \nabla \cdot {\Vec{q}} = 0 
    \\
    \label{eq3}
    \frac{\partial c}{\partial t} = -\frac{\Vec{q}}{\phi}\cdot \nabla c + \nabla \cdot (D\nabla c) 
\end{gather}

where $\Vec{q}$ is the Darcy flux, $k(x)$ is the log normally distributed random permeability field, $\mu(c)$, $\rho(c)$ is the concentration dependent viscosity and density of the miscible phase respectively, $\nabla P$ represents the pressure gradient, $g$ is the acceleration due to gravity, $\Vec{e}_z$ is a unit-vector pointing in the direction of positive y-axis, $c$ is the $\mathrm{CO_2}$ concentration, $\phi$ is the local porosity and $D$ is the hydrodynamic dispersion tensor represented as \citep{bear2013dynamics}:

\begin{gather}
\label{eq4}
D = (\alpha_T|q|+D_m)\delta + (\alpha_L-\alpha_T)\frac{qq^T}{|q|}
\end{gather}

$\alpha_L$, $\alpha_T$ are the longitudinal and transverse dispersivities respectively, $D_m$ is the mass diffusion coefficient and $\delta$ is the Kronecker Delta represented by an identity matrix. 

Here we consider the simplified form of the continuity equation, represented in equation \ref{eq2}, after applying the Boussinesq approximation, which takes into account only the influence of the specific weight of the fluid. Equation \ref{eq3}, commonly known as the advection-dispersion equation \citep{van2016simulation}, represents the transport of a passive scalar quantity under the influence of an ambient flow field. The non-linearity is imparted by the dependence of the Darcy flux on the local solute concentration as shown in equation \ref{eq1}. 

Our flow and transport solver incorporates all necessary attributes that are essential for modelling the above process. We adopt an operator splitting procedure to linearize the problem and use two distinct methods --- an Eulerian approach based on Finite Volume Method (FVM) to solve the flow field and a Lagrangian random walk particle tracking method (RWPT) to solve for the transport of the $\mathrm{CO_2}$--brine solute mixture. The linearization approach works in two steps --- solving for the Darcy flux to obtain the pore-water velocity and using this velocity field to advect particles subsequently. Overall, the module structure of PyDDC comprises of three major blocks that efficiently interacts to solve the coupled flow and transport problem and are described below:

\begin{itemize}
    \item Flow module - Solves the non-linear Darcy flux in a 2D Cartesian grid using FVM. 
    \item Phase module - Computes the different thermodynamic properties like $\mathrm{CO_2}$ solubility, density,  viscosity and diffusion coefficient of the $\mathrm{CO_2}$--brine mixture at a specified temperature and pressure.  
    \item Transport module - Solves the Lagrangian particle dynamics using RWPT approach and estimates concentration required by the flow module at specific set of points through a mapping between the Lagrangian point masses and Eulerian set of nodes. 

\end{itemize}

\begin{figure}[h]
    \centering
    \includegraphics[width=8 cm]{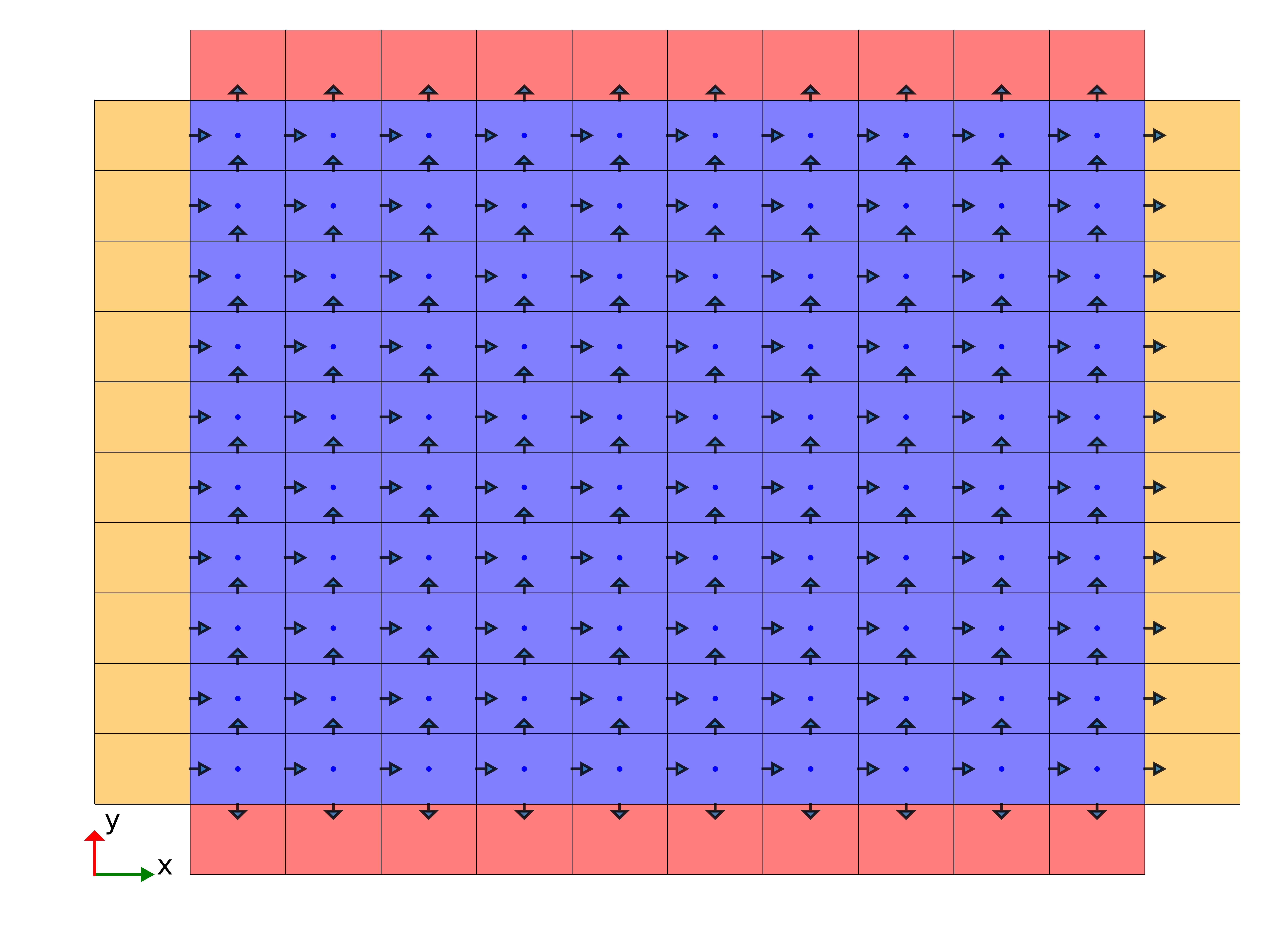}
    
    \caption{2D computational domain indicating scalar and vector quadrature points or nodes where discrete values of those variables are computed and stored. Quadrilateral cells are of 3 types, blue --- interior cells (domain of interest), orange and red --- ghost cells (introduced to enforce boundary conditions, as described in the text). Blue dots represent nodes where all scalar fields along with phase attributes are stored --- pressure, $k-\phi$ field, density, concentration, viscosity, diffusion coefficient. Black arrows at facets indicate nodes where all vector quantities are stores, namely the flux and scalar field gradients.}
\label{fig:1}
\end{figure}

\begin{figure}[t]
	\centering
 \includegraphics[width=8 cm]{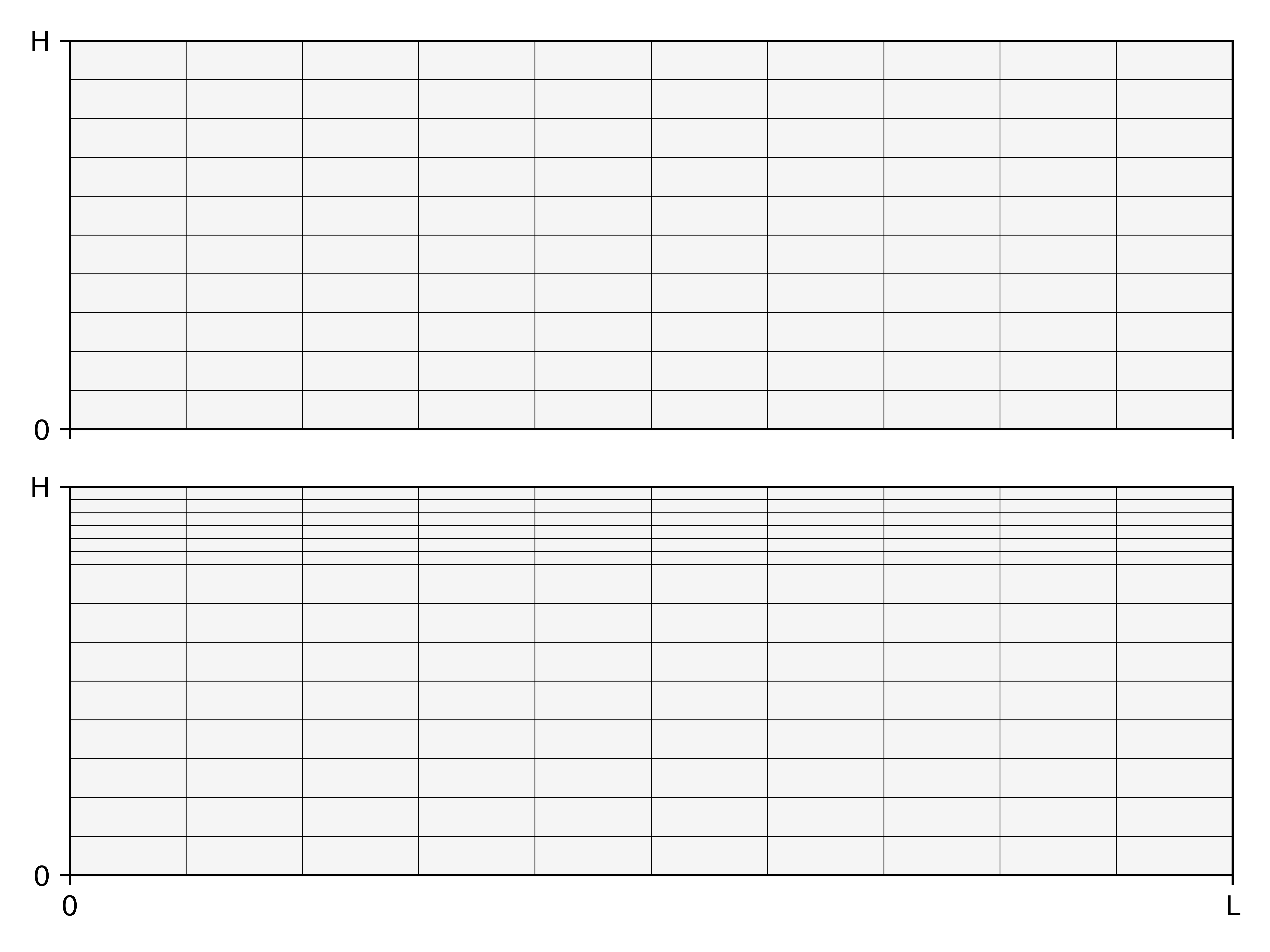}
 \caption{Grid refinement supported by PyDDC showing the unrefined regular mesh (top) and refined mesh (bottom).}
\label{fig.2}
\end{figure}

Currently PyDDC only supports orthogonal structured meshes and has the capability to handle unidirectional grid refinement along the vertical axis as demonstrated in Fig. \ref{fig.2}. This can be considered to be adequate as the primary objective behind refinement is to capture the transition from diffusive to advective scales, i.e. the onset of convection, with reasonable accuracy. In Fig. \ref{fig:1} we introduce the computational domain used by PyDDC to solve for the pressure field and Darcy flux. Here we adopt a staggered grid arrangement owing to its better conservation properties and its natural ability to directly enforce pressure-velocity coupling \citep{piller2004finite}. This approach also allows for the direct enforcement of free slip conditions on the fluxes at the boundaries of the domain. Two different types of ghost cells are indicated for their different usage. The red ghost cells at the top and bottom corresponds to impermeable layers, with the former being also used for applying constant concentration boundary condition. The left and right ghost cells, marked with orange, are used for applying the pressure boundary conditions to get a prescribed background flow.

\subsection{Flow Solver}
As shown in Fig. \ref{fig:1} the entire domain, $\mathrm{\Omega}$, is discretized into a finite number of cells, $\mathrm{N_x}$ and $\mathrm{N_y}$, in both x and y directions. Considering non-uniformity of the grid, a generic representation of cell centered co-ordinate, $x_c$, $y_c$, is given as:

\begin{equation}
\begin{aligned}
\label{eq5}
x_c=\frac{x_{i+1}+{x_i}}{2}(i=\mathrm{0,..,N_x}) \\
y_c= \frac{y_{j+1}+{y_j}}{2}(j=\mathrm{0,..,N_y})
\end{aligned}
\end{equation}

with $x_i\mathrm{'s}$ and $y_j\mathrm{'s}$ representing the cell boundaries and $\Delta x_i=x_{i+1}-{x_i}$ and $\Delta y_j=y_{j+1}-{y_j}$ representing the cell widths. 

\begin{figure}[t]
	\centering
    \includegraphics[width=\linewidth, height=11 cm]{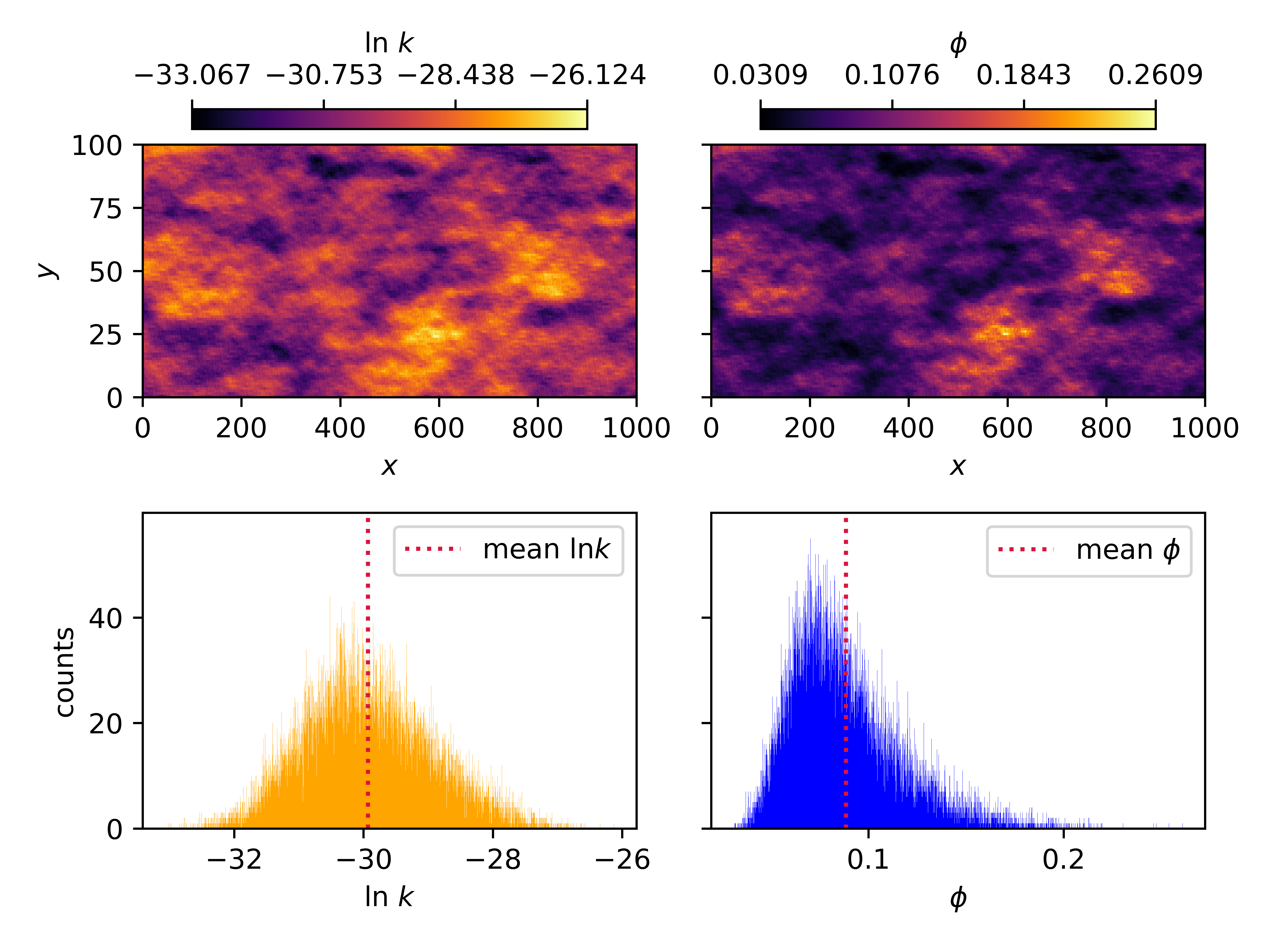}
\caption{Distribution of log normal permeability and porosity fields and their corresponding density plots showing the distribution around mean.}
\label{fig.3}
\end{figure}

The GSTools library developed by \cite{muller2022gstools} is used here to generate spatial random permeability fields with a specified covariance matrix, mean permeability, log-variance and axial correlation lengths. This package is employed into our flow module to obtain the Darcy flux and pressure field for a heterogeneous domain with a given set of user defined meta-parameters. The porosity field is determined from the permeability field via the Kozeny-Carman relation \citep{pape2000variation} $k(x) = \frac{\phi(x)^3}{c(1-\phi(x))^2S^2}$, where $\phi(x)$ is the porosity field, $c$ is the Kozeny constant and $S$ is a parameter depending on the grain size of the media. The realization of the random porosity-permeability ($k-\phi$) field along with their respective distributions are shown in Fig. \ref{fig.3}.

For the FVM implementation, the discretized equations subjected to necessary boundary conditions are formulated as:

\begin{equation}
\begin{aligned}
\label{eq6}
    \Vec{u}_{i+1/2,j} &= -\frac{k_{i+1/2,j}}{\mu_{i+1/2,j}}\frac{\partial P_{i+1/2,j}}{\partial x}  &\mathrm{in} \hspace{0.25 cm} \Omega\\
    \Vec{v}_{i,j+1/2} &= -\frac{k_{i,j+1/2}}{\mu_{i,j+1/2}}(\frac{\partial P_{i,j+1/2}}{\partial y} + \rho_{i,j+1/2} g \Vec{e}_z) &\mathrm{in} \hspace{0.25 cm} \Omega\\
    \frac{\partial u}{\partial x} + \frac{\partial v}{\partial y} &= 0 &\mathrm{in} \hspace{0.25 cm}\Omega \\
    \Vec{v}_{i,\mathrm{H}}\cdot n &= 0 &\mathrm{on} \hspace{0.25 cm}\delta \Omega_N \\ 
    \Vec{v}_{i,\mathrm{0}}\cdot n &= 0 & \mathrm{on} \hspace{0.25 cm}\delta \Omega_N \\
    P_{i-1/2,j} &= (\mathrm{P_{top}}+\mathrm{P_{il}}) \rho_{i,j} gy_c &  \mathrm{on} \hspace{0.25 cm}\delta \Omega_D \\
    P_{i+1/2,j} &= (\mathrm{P_{top}}+\mathrm{P_{ir}}) \rho_{i,j} gy_c &  \mathrm{on} \hspace{0.25 cm} \delta \Omega_D
\end{aligned}              
\end{equation}

As the Darcy flux is entirely dictated by the pressure gradient and density term, we first solve for the pressure Poisson equation and use this computed pressure field to solve for the flux. Substituting equation \ref{eq2} in equation \ref{eq1} and integrating over the domain, we get:

\begin{align}
\label{eq7}
\int_{\Omega} \nabla \cdot \Vec{q} dV &= \int_{\Omega} \nabla \cdot (-\frac{k(x)}{\mu(c)}(\nabla P + \rho(c) g \Vec{e}_z)) dV = \int_{S} {-\frac{k(x)}{\mu(c)}(\nabla P + \rho(c) g \Vec{e}_z)} \cdot n dS = 0
\end{align}
where the above transformation from volume integral to surface integral is attained with the help of Green-Gauss theorem.
This equation can be discretized as summing over all the facet elements belonging to each cell:
\begin{align}
\label{eq8}
    \Sigma_{f} -\frac{k}{\mu}(\nabla P + \rho g \Vec{e}_z) \vert _f \cdot n_f\Delta S = 0 
\end{align}
 The boundary conditions on the pressure and the fluxes are represented in equation \ref{eq6}, where the subscript $D$ refers to Dirichlet boundary and subscript $N$ denotes Neumann boundary. The computational domain along with the different boundary conditions are highlighted in Fig. \ref{fig.3}. Currently we only support imposition of one type of boundary condition, which is considered to be the most natural case for this simulation. Free slip boundary conditions are imposed on the Darcy flux at the top and bottom boundaries, and the pressure is taken to be hydro-static at the left and right boundaries. We consider the most common form of pressure gradient in an aquifer, represented as $P=\mathrm{P_t} + \rho g h$, where $\rho g h$ corresponds to the hydrostatic term and $p_t$ being the pressure at the top of the aquifer. The background flow into the domain can be introduced by providing an increment of pressure ($\mathrm{P_{il}}$ or $\mathrm{P_{ir}}$) at any boundary on $\Omega_D$. 

 These two fields can either be specified by the user prior to model execution or can be initialized by the inbuilt random field generator. 
  After obtaining the random $k-\phi$ field and knowing the value of the viscosity and density (obtained either from phase module or from predefined user specified values), the flow equation can be solved to obtain the pressure and Darcy flux.

\begin{figure}[h!]
\centering
	     \includegraphics[width=\linewidth]{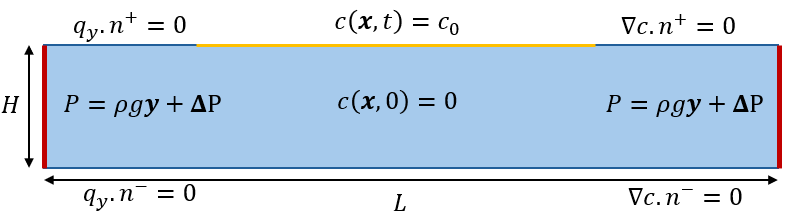}
    \caption{Domain representation showing different boundary conditions. Yellow line at the top indicates the equilibrium concentration corresponding to the solubility of $\mathrm{CO_2}$ at the interface, the length of which corresponds to the lateral extent of source. Vertical red lines at both ends represents the hydrostatic pressure boundary condition. Boundary constraints on the velocity and the concentration are marked at the top and bottom. $n^+$ and $n^-$ represents the normal components perpendicular to the top and bottom surface pointing outwards. $q_y\cdot n=0$ and $\nabla c\cdot n=0$ indicates zero fluxes for both the scalar as well as the vector fields across the boundary surface.}
    \label{fig.4}
\end{figure}

To construct the global linear system from equation \ref{eq7} for both uniform and non-uniform grids, we adopt the 2nd order accurate scheme (Scheme-II of 
\cite{wangcentral}). We introduce new indexing notation based on the five point stencil, where we denote the center cell $ij$ by $c$ and the neighbouring cells as $n$, $e$, $s$, $w$ corresponding to ($i,j+1$), ($i+1,j$), ($i,j-1$), ($i-1,j$) cell indices. We define weights, $w$'s, used for linear interpolation from cell-centers to cell-interfaces which are given by the volume ratio between the current cell and the neighbouring cells according to:

\begin{equation}
\begin{aligned}
\label{eq9}
    w_n &= \frac{\Delta V_{i,j}}{\Delta V_{i,j+1} + \Delta V_{i,j}} \\
    w_s &= \frac{\Delta V_{i,j}}{\Delta V_{i,j-1} + \Delta V_{i,j}} \\
    w_e &= \frac{\Delta V_{i,j}}{\Delta V_{i+1,j} + \Delta V_{i,j}} \\
    w_w &= \frac{\Delta V_{i,j}}{\Delta V_{i-1,j} + \Delta V_{i,j}}   
\end{aligned}
\end{equation}

Following this and assembling the terms on the left and right hand side, we get the final form of our pressure equation:

\begin{align}
    \label{eq10}
    (L_nP_n + L_cP_c + L_sP_s) + (L_eP_e + L_cP_c + L_wP_w) &= R_s\rho_s + R_c\rho_c + R_n\rho_n
\end{align}
where, 
\begin{equation}
\begin{aligned}
\label{eq11}
    R_s &= k_{i,j-1/2}w_sg\Delta x_{i,j} \\
    R_c &= (k_{i,j-1/2}(1-w_s) - k_{i,j+1/2}(1-w_n))g\Delta x_{i,j} \\
    R_n &= - k_{i,j+1/2}w_ng\Delta x_{i,j} \\
    L_c &= L_n - L_s + L_e - L_w \\
    L_n &= \frac{\Delta x_{i,j}}{\Delta y_{i,j} + \Delta y_{i,j+1}}\biggl[\frac{\Delta y_{i,j}}{\Delta y_{i,j+1}} - w_n\left(\frac{\Delta y_{i,j}}{\Delta y_{i,j+1}} - \frac{\Delta y_{i,j+1}}{\Delta y_{i,j}}\right)\biggr] k_{i,j+1/2} \\
    L_s &= \frac{\Delta x_{i,j}}{\Delta y_{i,j} + \Delta y_{i,j-1}}\biggl[\frac{\Delta y_{i,j}}{\Delta y_{i,j-1}} - w_s\left(\frac{\Delta y_{i,j}}{\Delta y_{i,j-1}} - \frac{\Delta y_{i,j-1}}{\Delta y_{i,j}}\right)\biggr]k_{i,j-1/2} \\
    L_e &= \frac{\Delta y_{i,j}}{\Delta x_{i,j} + \Delta x_{i+1,j}}\biggl[\frac{\Delta x_{i,j}}{\Delta x_{i+1,j}} - w_e\left(\frac{\Delta x_{i,j}}{\Delta x_{i+1,j}} - \frac{\Delta x_{i+1,j}}{\Delta x_{i,j}}\right)\biggr]k_{i+1/2,j} \\
    L_w &= \frac{\Delta y_{i,j}}{\Delta x_{i,j} + \Delta x_{i-1,j}}\biggl[\frac{\Delta x_{i,j}}{\Delta x_{i-1,j}} - w_w\left(\frac{\Delta x_{i,j}}{\Delta x_{i-1,j}} - \frac{\Delta x_{i-1,j}}{\Delta x_{i,j}}\right)\biggr]k_{i-1/2,j} 
\end{aligned}
\end{equation}

The $P$'s are the unknown pressure terms and $\rho$ is the density of the mixture which is a function of the local concentration.

\begin{figure}[t]
    \centering
    \includegraphics{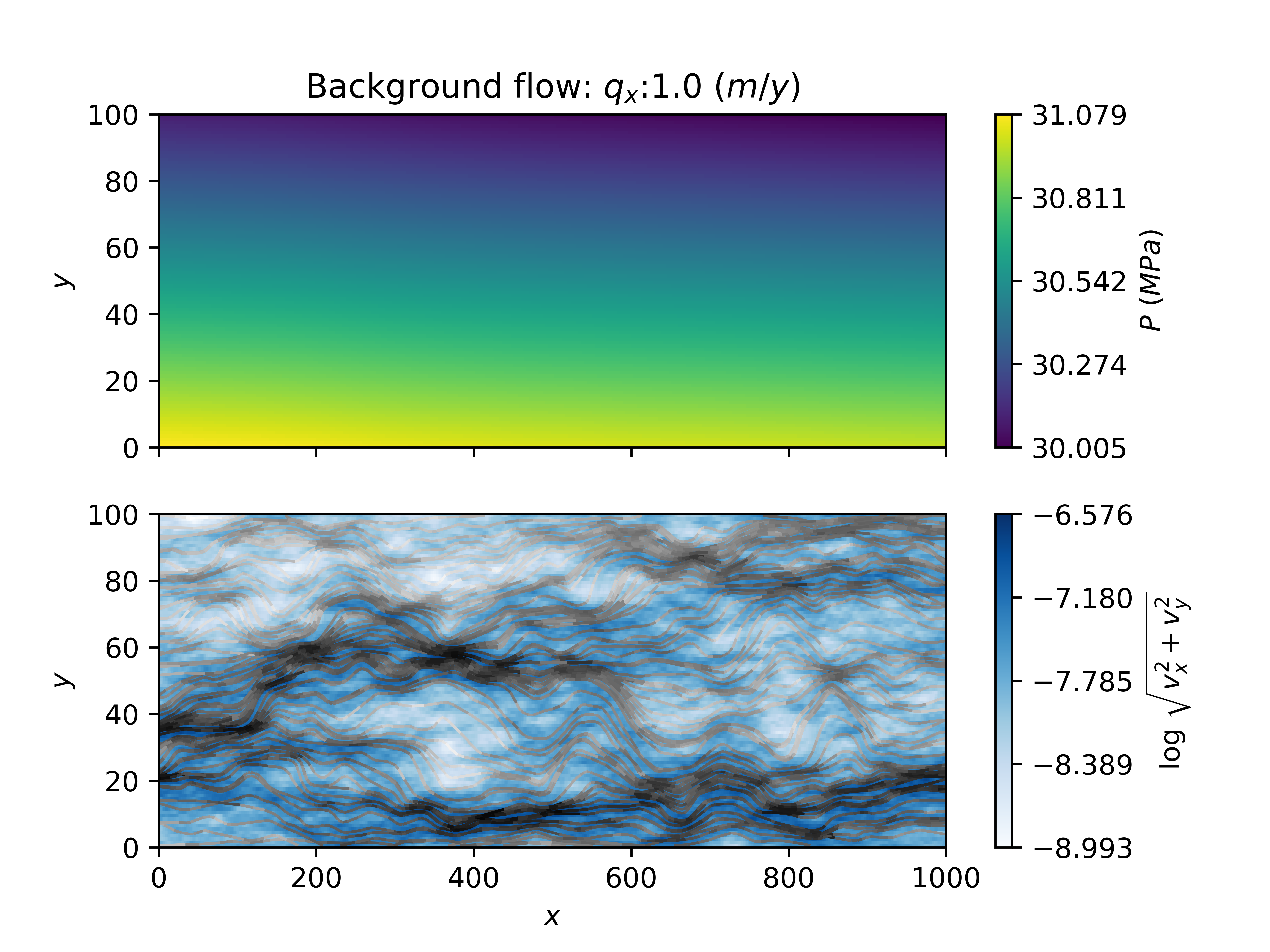}
    \caption{Initial condition configured from a predefined global temperature and pressure values.}
    \label{fig:5}
\end{figure}
Assembling equation \ref{eq9} results in the linear system of equations in the form: $$A\Vec{x}=\Vec{b}$$ where, $A$ is the coefficient matrix, $x$ is the vector of unknowns and $b$ is the vector of known coefficients. 
For optimization purpose, we take special care of solving the linear system of equations. If the fluid viscosity is constant, the pressure coefficient matrix $A$ is invariant. So, we perform a LU decomposition on A and precompute its inverse, $A^{-1}=(LU)^{-1}$, only once at the start of the simulation to be used at successive iterations. On the other hand, if the viscosity is concentration dependent, then the pressure coefficient matrix cannot be precomputed. Hence, in this case, we use a sparse linear solver using the previous iteration pressure values as the initial guess for faster convergence. This approach proves to be really effective even if we are dealing with very high medium heterogeneity.

To solve the non-linear Darcy equation (\ref{eq1}) requires us to know the density field locally, i.e. cellwise. To estimate any phase attribute at any point in space, we should know the pressure and temperature value associated with that point. For our case we have an isothermal model with a global predefined temperature and pressure which is insufficient to get a density field.  So we configure an initial condition for our problem assuming the local pressure everywhere to be same as the pressure at the top and based on this we can get the density field which can then be used to solve for the pressure field which in turn can be used in successive iterations. Figure \ref{fig:5} shows one such generated initial condition from a predefined global pressure of $30$ MPa and temperature of $50$ $^{0}$C.

\subsection{Phase Equilibrium Model}

 In order to solve the flow and transport equations, it is imperative to estimate the phase attributes of the $\mathrm{CO_2}$--brine mixture, namely the solubility, density, viscosity and $\mathrm{CO_2}$ diffusivity at the ambient pressure and temperature.  This is done with the help of the thermodynamic models that are developed extensively over the past years in the context of $\mathrm{CO_2}$ storage at geological conditions. The associated complexity of estimating all the four above mentioned phase parameters demands a detailed description of each individually. 

 \subsubsection{Solubility Model}
 
 We use the model of \cite{sun2021simple} for estimating $\mathrm{CO_2}$ solubility in brine in a non-iterative manner. Although their model is capable of estimating water content in the $\mathrm{CO_2}$ rich phase, for our case we neglect it and only focus on the solubility of $\mathrm{CO_2}$ in the aqueous phase. We refer to the procedure as presented in \cite{sun2021simple} where first the $\mathrm{CO_2}$ solubility in pure water is computed and then the necessary changes for the presence of electrolyte are taken into account in the activity coefficient. This model is hence faster and more computationally efficient compared to other popular thermodynamic models like the Spycher EoS \citep{hassanzadeh2008predicting}. 
 
 \subsubsection{Density Model}
 \label{2.2.2}
 The density model for ternary $\mathrm{CO_2}$-$\mathrm{H_2O}$-NaCl system, developed by \cite{duan2008densities}, is given as:
\begin{align}
\label{eq12}
\rho &= \frac{x_1M_1 + x_2M_2 + x_3M_3}{V} \\
\label{eq13}
V &= x_1V_1 + x_2V^B_{\phi,2} + x_3V^B_{\phi,3}
\end{align}
where $\rho$ is the solution density, $V$ is the solution volume, $x_1$,$x_2$, $x_3$ are the mole fractions of $\mathrm{H_2O}$, NaCl and $\mathrm{CO_2}$, with $M_1$, $M_3$, $M_3$ being the corresponding molecular weights. $V_1$ is the molar volume of $\mathrm{H_2O}$, and $V^B_{\phi,2}$, $V^B_{\phi,3}$ represents the apparent molar volume of electrolyte and $\mathrm{CO_2}$ in $\mathrm{H_2O}$ respectively. The apparent molar volume of $\mathrm{CO_2}$ in $\mathrm{H_2O}$ and brine is considered to be equivalent due to the very low solubility of $\mathrm{CO_2}$ in brine at geological storage conditions. We use the empirical model of 
\cite{hu2007pvtx} to compute the molar volume of $\mathrm{H_2O}$. 
We use the well known Pitzer model \citep{rogers1982volumetric} to first compute the apparent molar volume of electrolyte, given by the relation:
\begin{align}
\label{eq14}
    V_\phi = V_\phi^0 + \nu|z_Mz_X|A_vh(I(m)) + 2\nu_M\nu_XRT(mB^V_{MX} + m^2v_Mz_MC^V_{MX})
\end{align}
where subscripts $M$, $N$ indicates the cation and anion species respectively, $\nu_M$, $\nu_N$ refers to the stoichiometric proportion of the respective ionic species, $z_M$, $z_N$ are their corresponding ionic charges, $R$ is the molar gas constant, $T$ is the temperature and $h(I(m))$ is a parameter depending on the ionic strength, $I(m)$, at a specified molality ($m$) and $B^V_{MX}$ and $C^V_{MX}$ are empirical parameters fitted to the model based on existing volumetric data.  $V_\phi$ and $V_\phi^0$ indicates the apparent molar volume and apparent molar volume at infinite dilution respectively. The apparent molar volume at infinite dilution is obtained from the same equation corresponding to the reference molality, $m_r$. From the relation between solution volume, $V(m)$, and apparent molar volume, $V_\phi$, given as: $$V_\phi=\frac{V(m)-\frac{1000}{\rho_{H_2O}}}{m}$$
at the reference molality, we get the expression for $V^0_\phi$:
\begin{align}
\label{eq15}
    V_\phi^0= \frac{V(m_r)}{m_r} + \frac{1000}{m_r\rho_{H_2O}} + \nu|z_Mz_X|A_vh(I(m_r)) + 2\nu_M\nu_XRT(m_rB^V_{MX} + m_r^2v_Mz_MC^V_{MX})
\end{align}
where $V(m_r)$ is also a fitted empirical parameter and $\rho_{H_2O}$ is the density of water computed from the IAPWS-97 model of \cite{wagner2008iapws}. After computing  $V^0_\phi$, we substitute it in equation \ref{eq14} to compute the apparent molar volume of the electrolyte brine binary mixture. 

This forms the basis of the density model which we invoke at each iteration to compute the density field of $\mathrm{CO_2}$--brine mixture based on the local variations of concentration and pressure. Previous studies have mainly relied on a linear variation of density with concentration considering the field to be monotonous, given by the simple relation:

\begin{align}
\label{eq16}
\rho(c) = \rho_w + \Delta\rho\frac{c}{c^0}
\end{align}

where $\Delta\rho = \rho_s-\rho_w$ is the density difference between the $\mathrm{CO_2}$ saturated brine ($\rho_s$) and  pure brine ($\rho_w$), $c$ is the local concentration and $c^0$ is the saturated concentration at the boundary. As can be seen, instead of computing the local density based on the proposed model, we can also compute the end member densities and use the above linear expression to find the values globally at any point. In fact we decide to keep both options available for the user to decide. 

\subsubsection{Viscosity Model}

\cite{sun2022modeling} extended the Goldsack-Franchetto model to incorporate $\mathrm{CO_2}$ as an additional electrolyte species based on the temperature and individual mole fractions, which is employed here to solve for the viscosity of mixture . For a mixed electrolyte system, comprising here of the two species NaCl and $\mathrm{CO_2}$, the relative viscosity is computed as:
\begin{align}
\label{eq17}
\eta_r = \exp(\Sigma_{i=1}^n X_iE_i) / (1 + \Sigma_{i=1}^nX_iV_i)
\end{align}
where, $E_i$ and $V_i$ are temperature dependent empirical parameters  given by:
\begin{align*}
E_i &= \Vec{C_1} \cdot \Vec{T} \\
V_i &= \Vec{C_2} \cdot \Vec{T} 
\end{align*}
where $C_1$, $C_2$ are vector of coefficients derived from model fitting and $T$ is the vector of polynomials represented as $\Vec{T} = [1, T, T^2]$. $X_i$ is the mole fraction of species $i$. 
\begin{align}
\label{eq18}
X_i = m_i / (55.51 + \Sigma_{i=1}^n\nu_im_i)
\end{align}
where $\nu_i$ are the stoichiometric coefficients of the ions and $m_i$ are their respective ionic molalities. For NaCl we consider complete dissociation to its respective ions, $Na^+$ and $Cl^-$ leaving with $\nu_i$ and $m_i$ of 1 each. For $\mathrm{CO_2}$, we consider weak dissociation, $CO_2 \rightarrow H^+ + HCO_3^-$, corresponding to a weak degree of ionization, giving a $\nu_i$ and $m_i$ also equal to 1. If the end member values of the viscosities are known the viscosity variation with concentration can be represented by an exponential relation in the form \citep{chen2023density}:
\begin{gather}
\label{eq19}
    \mu = \mu_0e^{R(c-c_0)}
\end{gather}
where $\mu$ is the dynamic viscosity,  $\mu_0$ is the dynamic viscosity at saturated concentration,  $c$ is the local concentration, $c_0$ is the concentration of $\mathrm{CO_2}$ at the Sc-$\mathrm{CO_2}$--brine interface and $R=\log (\frac{\mu_{CO2-br}}{\mu_{br}})$ is the mobility ratio with $\mu_{CO2-br}$ and $\mu_{br}$ being the end-member dynamic viscosities. Whether to use this or the 
Goldsack-Franchetto model lies entirely in the discretion of the user.  

\subsubsection{Diffusivity Model}

From molecular dynamics simulation studies, \cite{omrani2022insights} proposed a generalized model for estimating $\mathrm{CO_2}$ diffusivity into brine based on temperature and salinity. Their model involves no pressure terms as the effect temperature and salinity are found to be more profound compared to pressure. We use their empirical relation:
 \begin{align}
\label{eq20}
D = -18.1579e^{-0.0574\hspace{0.08 cm}C} + 0.0687\hspace{0.08 cm}T - 0.0004\hspace{0.08cm}C^{0.8206}\hspace{0.08cm}T^{1.4633}
\end{align}
where $D$ is the diffusion coefficient of $\mathrm{CO_2}$ in $\mathrm{m^2}$, $C$ is the concentration of electrolyte represented on the molarity scale ($\mathrm{mol/litre}$) and T is the temperature in $^0\mathrm{K}$. All the coefficients in equation \ref{eq18} are rounded to four decimal places.

Each individual model is applicable over a wide range of temperature and pressure covering the whole range of $\mathrm{CO_2}$ sequestration regime up to an electrolyte molality of 6 mol/kg. Although we have incorporated the $\mathrm{CO_2}$-$\mathrm{H_2O}$-NaCl thermodynamic model, we keep the option available for the user to not use our phase module. In this case the user needs to feed the values of the phase parameters to the flow solver at each iteration which would then be used to run the simulation.

\subsection{Lagrangian Particle Tracker}
 
A particle tracker, as its name implies, is a Lagrangian approach that inherently tracks discrete particles representing some physical quantity. For our case that physical quantity being tracked is the $\mathrm{CO_2}$--brine mixture concentration. The tracks of those individual particles are stochastic in nature and are governed by the generalized stochastic differential equation (GSDE) as given by the integration scheme \citep{salamon2006review}:
\begin{align}
\label{eq21}
X_p(t+\Delta t) = X_p(t) + \frac{1}{\phi}A(X_p,t)\Delta t + B(X_p,t)\cdot \zeta(t)\sqrt{\Delta t}
\end{align}
where $dX = X_p(t+\Delta t) - X_p(t)$ represents the displacement of the particle in an interval $\Delta t$, $A(X_p,t)=v(X_p,t)$ is the deterministic drift vector, $B(X_p,t)$ is the displacement matrix and $\zeta(t)\sim \mathcal{N}(0, 1)$ is a white noise parameter. The only caveat is that equation \ref{eq21} solves the Fokker-Planck equation and not the advection-dispersion equation (ADE) represented in equation \ref{eq3}. The equivalence between the ADE and Fokker-Planck lies in rearranging the terms in ADE such that we end up with an augmented drift vector $A(X_p,t)=(v + \phi \nabla\cdot D + \frac{1}{\phi}D\cdot\nabla\phi)(X_p, t)$ for a heterogeneous case \citep{henri2022unsaturated}. For a homogeneous case we drop the tailing term and end up with the two leading terms $v$ and $\nabla\cdot D$. For an isotropic case in 2D, the displacement matrix, $B(X_p, t)$, is represented as \citep{salamon2006review}:
\begin{align}
\label{eq22}
B(X_p, t) &=
\begin{bmatrix}
   \frac{u_x}{|u|}\sqrt{2\alpha_L|u|}       & -\frac{u_y}{|u|}\sqrt{2\alpha_L|u|}  \\
    \frac{u_y}{|u|}\sqrt{2\alpha_L|u|}    &  -\frac{u_x}{|u|}\sqrt{2\alpha_L|u|} \\
\end{bmatrix}
\end{align}

where $\alpha_L$ is the longitudinal dispersion coefficient, $u_x$ and $u_y$ are the horizontal and vertical velocity components respectively and $|u|$ is the Euclidean norm of the vector field $u$. 
This aforementioned scheme is also popularly known as random walk particle tracking (RWPT). Although RWPT methods are easier to implement to track individual particle trajectories, additional care must me taken such that the estimated density respects mass conservation and boundary conditions. As the particles we consider are point masses that are dispersed in the continuum, it does not possess a metric to compute field values like concentration. This makes the concentration estimated from particle locations to be a delta function at the Lagrangian node where the particle resides and is given as: 
\begin{align}
\label{eq23}
c(x,t) = \Sigma_{p\in N_p}m_p(t)\delta(x-X_p(t))
\end{align}
where $c(x,t)$ is the spatially varying concentration field, $x$ is any point within the domain. But to solve the non-linear problem we need to estimate the concentration values at the nodes where the flow velocity is estimated and therefore an accurate mapping from Lagrangian nodes to Eulerian nodes is required. For this reason, the metric to be estimated, i.e. the mass of the particles,  are considered to be distributed symmetrically or asymmetrically around each particle giving rise to a particle support volume. This is facilitated by using a projection function in the form of a kernel that acts as an operator to estimate the concentration at any point in the domain and is represented as \citep{bagtzoglou1992projection}:

\begin{align}
\label{eq24}
\Tilde{c}(x,t) = \int_\Omega \hat{c}(x',t)W(x-x')dx' = \frac{1}{\phi(x)}\Sigma_{p\in N_p}m_p(t)W(x-X_p(t))
\end{align}

where $W(x-X_p(t))$ is the kernel having a support volume whose evaluation is a function of the position of the particle, $X_p$ and the point of estimation, $x$, within a predefined interval governed by the kernel bandwidth. This method is also popularly known as kernel smoothing as we find a smoothed continuous distribution of a field from its discrete representation. The kernel contains a bandwidth parameter, $h$, responsible for the quality of density estimation and can be either considered to be a globally uniform value or a locally adaptive one \citep{sole2018lagrangian}. In order to be a valid approximation of the true concentration field and ensure mass conservation, the kernel function should satisfy the condition that $\int_\Omega W(x-X_p(t))dx=1$, where $\Omega$ is the domain of interest. For simplicity and computational efficiency we choose the standard simple binning method with the support volume being the dimensions of the grid cells. This makes the estimated concentration, $\hat{c}=\frac{N_p}{\phi\Delta V}$ which is basically the number of particle in each grid cell divided by the available cell volume \citep{wright2018upscaling}. 

\begin{figure}[h]
	\centering
	     \includegraphics{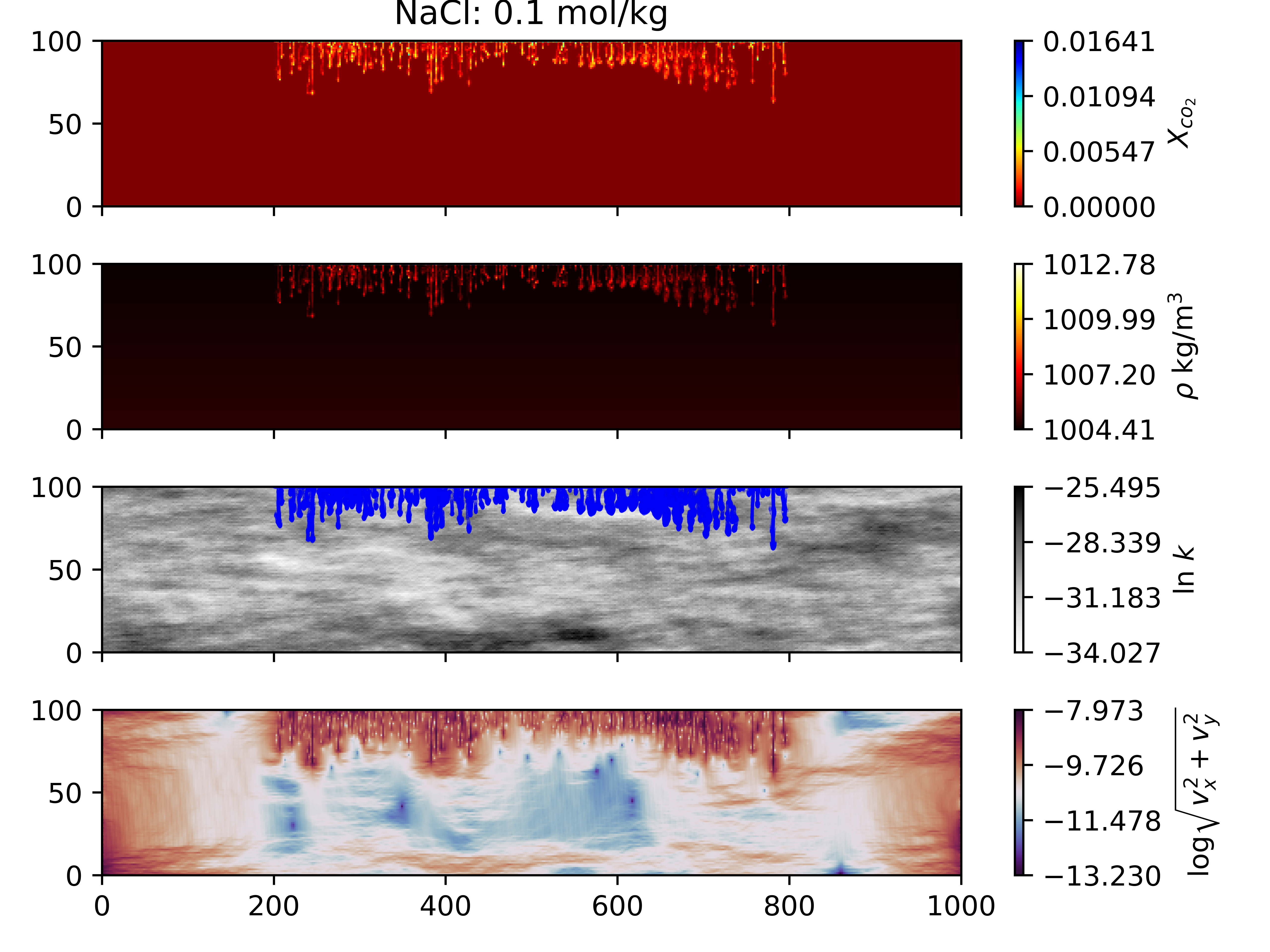}
      \caption{Fingering stage for a heterogeneous case at $t=53$ years for an ionic molality of $0.1 \mathrm{mol/kg}$ showing the concentration field, density field, particle configuration and velocity field (from top to bottom).}
      \label{fig:6}
\end{figure}
In addition to this, we also need to model the particle influx from the top boundary where we have a constant $\mathrm{CO_2}$ solubility computed at each iteration from the solubility model. It is important to note here that as the kernel is also locally conservative, we can enforce the Dirichlet boundary condition simply by reverse engineering. We divide the reservoir into discrete grid cells that acts as mirror reflections of the interior cells. After the binned concentration is computed at each step, based on the $\mathrm{CO_2}$ solubility at the boundary, we can define the reservoir concentration such that the boundary conditions are strictly satisfied. After estimating the concentration values, we create particle clouds inside each reservoir control volume which is basically the grid to particle projection, contrary to the particle to grid projection that we described previously. Now once we have those particle clouds created, we apply the simple Fickian diffusion physics to diffuse particle inside domain with a characteristic diffusion coefficient, $D$. The top boundary which the reservoir shares with the computational domain is held symmetric w.r.t diffusion in order to properly impose the diffusive physics, which means that we increase particle count as more particles diffuses from the reservoir to the domain and at the same time delete particles that leave the domain from the top boundary. But this boundary is closed w.r.t dispersion. As we have different boundary conditions for the diffusive and dispersive physics, we decompose the random walk algorithm by first diffusing particles, correcting particle count by applying diffusion boundary conditions and then apply the 2D random walk represented by equation \ref{eq21} respecting the dispersion boundary conditions.

Figure \ref{fig:6} shows the plume structure as represented by particle cloud and the corresponding density and concentration fields obtained after an elapsed simulation time of around 53 years. Addition to estimating concentration, density, pressure and velocity fields which are stored in binary format, HDF5, to be accessed and used by the user for different post-processing applications, we also store another important attribute, the vorticity field. In 2D, vorticity represents the magnitude of convection which can be used to understand the circulation inside an enclosed domain. For a 2-component vector field $\Vec{q}=[u,v]$, vorticity, $\omega$, is given by the relation: 

\begin{align}
\label{eq25}
\omega &= \curl \Vec{q} = \frac{\partial v}{\partial x} - \frac{\partial u}{\partial y}
\end{align}

For our case,

\begin{align}
\label{eq26}
\curl \Vec{q} &= \curl -\frac{k}{\mu}(\nabla P + \rho g\Vec{e_z}) = -\frac{k}{\mu}\frac{\partial \rho}{\partial x}g 
\end{align}
From the above expression it can be seen that once the fingers starts propagating, the convection cells starts to develop due to the horizontal density gradients between the fingers and its magnitude increases with increasing density finger propagation. The convective cells are indicated by the velocity plot in Fig. \ref{fig:6} showing the magnitude that advective velocities attain during convection.

\sloppy
\section{Using the software}
\subsection{Parameter specification through JSON file}

\begin{figure}[h]
    \centering
    \includegraphics[width=0.85\textwidth]{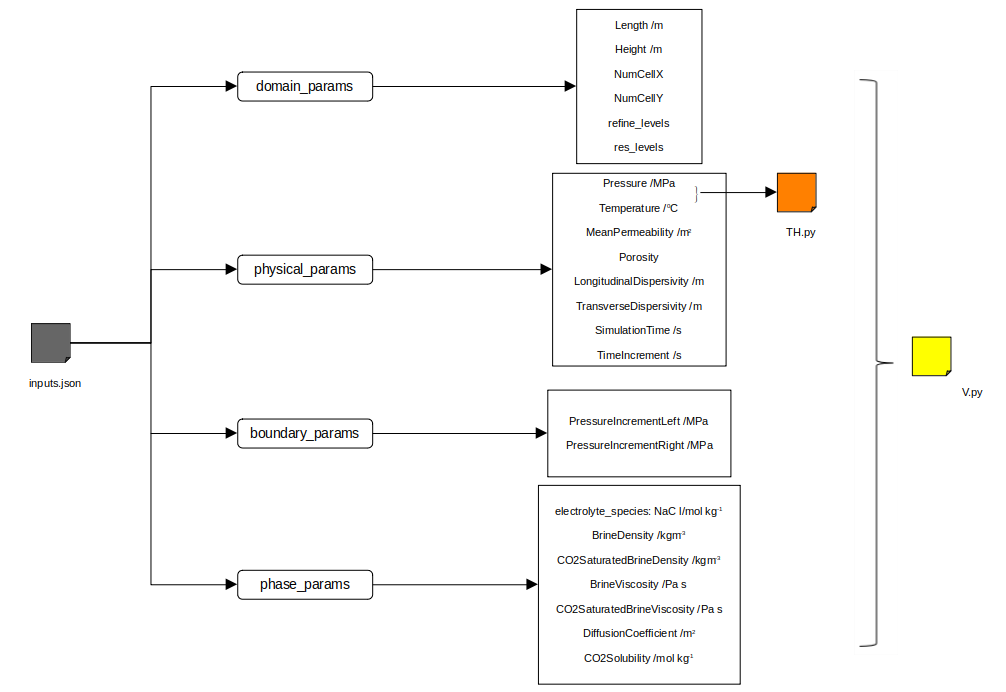}
    
    \caption{Internal structure of the JSON file showing the different input parameters and their corresponding units. The overall content is arranged in a hierarchical manner with different attributes grouped under their corresponding types for better readability. The Python files \texttt{V.py} and \texttt{TH.py} inherits parameters from the JSON file and converts them into Python objects to be accessed everywhere inside the software}
    \label{fig:9}
\end{figure}

The user input to the model is accepted through a JSON file present at the top level of the working directory, that holds the physical attributes required for the simulation. A JSON file represents structured data as key value pairs, with the keys holding the attributes names and the values being their corresponding quantification. Figure \ref{fig:9} gives the detailed insight into the JSON file with the physical parameters along with their units. All the physical parameters are grouped under four categories depending on the parameter type. \texttt{domain\textunderscore params} groups all the parameters required for defining the computational domain. Grid refinement is achieved by two input parameters --- \texttt{refine\textunderscore levels}, which takes into account the total number of cells from the top where refinement has to be applied and \texttt{res\textunderscore levels}, which holds the factor by which the initial resolution in the y-direction should be increased for cells affected by refinement. Transition zones due to refinement are handled internally in order to avoid considerable jumps in the cell aspect ratios between the refined and unrefined zones. \texttt{kfield\textunderscore params} holds the necessary meta-parameters, \texttt{mean\textunderscore k}, \texttt{var} and \texttt{corr\textunderscore length}, required to create the heterogeneous random permeability field. This field, \texttt{kf}, is created by invoking the  \texttt{Field} class function \texttt{Field.KField(x, y, mean\textunderscore k, var, corr\textunderscore length)} where \texttt{x} and \texttt{y} represents the spatial coordinates. For simulations requiring heterogeneous porosity fields, the \texttt{Field} class attribute \texttt{Field.PHIField(kf)} is used. \texttt{boundary\textunderscore params} contains the dirichlet pressure boundary conditions required for solving the Darcy flux. \texttt{reservoir\textunderscore params} contains the lateral extent of the Sc-$\mathrm{CO_2}$ source, \texttt{horizontal\textunderscore extent} and \texttt{molesperparticle}, which is a quantity carried by individual particles. The user-defined phase attributes for the model are stored in \texttt{phase\textunderscore params}, which contains the salinity (\texttt{NaCl}), saturated concentration (\texttt{CO2SaturatedConcentration}), $\mathrm{CO_2}$ diffusion coefficient in brine (\texttt{DiffusionCoefficient}) and end-member viscosity (\texttt{CO2SaturatedBrineViscosity}, \texttt{BrineViscosity}) and density (\texttt{CO2SaturatedBrineDensity}, \texttt{BrineDensity}). All user-specified attributes in the JSON file are converted to Python objects and are initialized in the Python file \texttt{V.py}. File \texttt{TH.py} holds the features required for computing phase attributes if the inbuilt phase module is used. These files are declared in the \texttt{formulate} module and can be  accessed, modified and used internally whenever required. 

\subsection{Workflow}

\begin{figure}[h]
    \centering
    \includegraphics[width=0.77\textwidth]{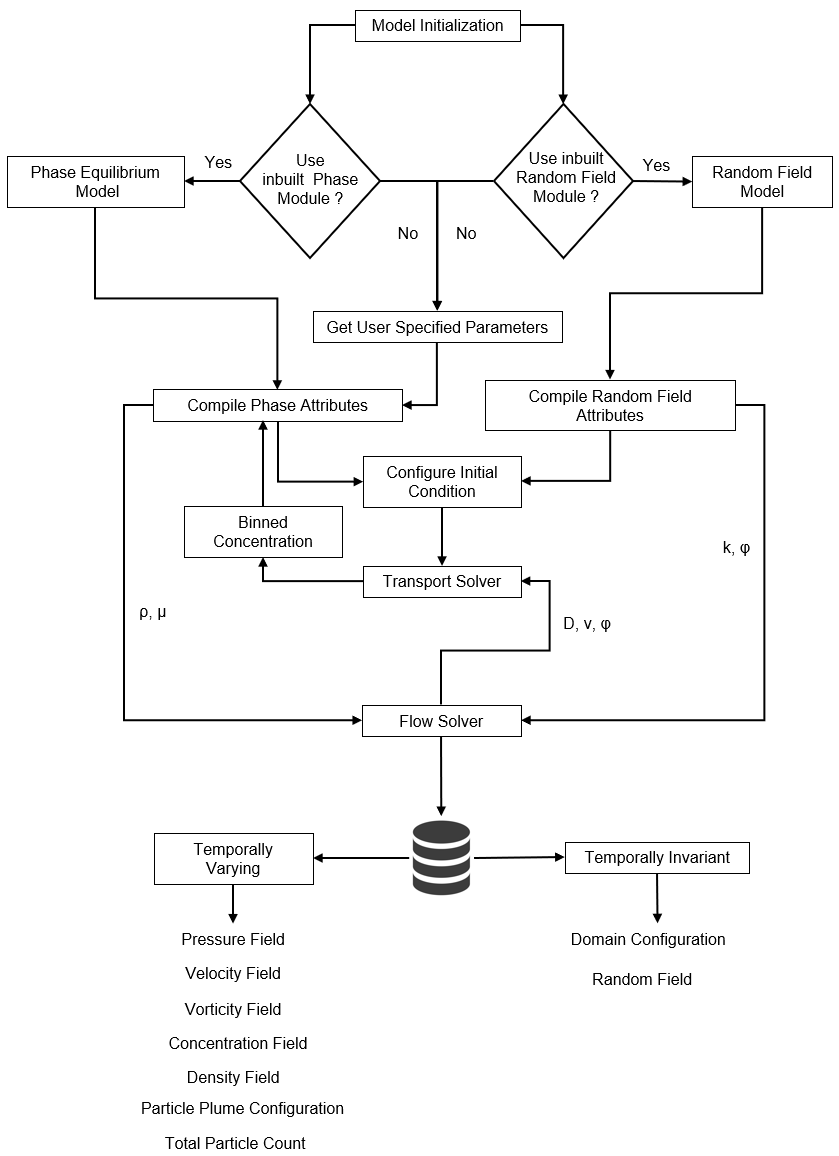}
    \caption{Flowchart showing the flow of control for the entire simulation process including the different module interactions.}
    \label{fig:10}
\end{figure}

\begin{figure}[t]
	\centering
 \includegraphics[width=0.8\linewidth]{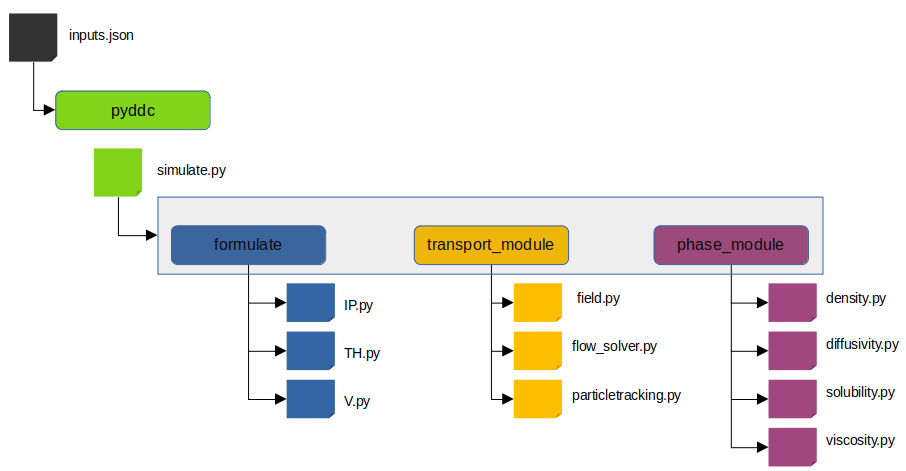}
 \caption{PyDDC software structure  showing the module arrangements and contents with the arrows indicating the hierarchy levels.}
\label{fig.11}
\end{figure}

The components of the numerical software along with their contents are shown in Fig. \ref{fig.11}. Proper execution of the numerical solver requires that the different software components interact efficiently with one another. This is leveraged by the systematic coupling of the different parts of the module as shown in Fig. \ref{fig:10}. The \texttt{Simulate} class handles the simulation process and is designed to establish the overall flow of control. To create a \texttt{Simulate} object, the user should pass the arguments --- \texttt{filename}, the JSON file which contains the physical parameters used in the simulation, \texttt{UsePhaseModule}, a boolean variable indicating whether to use the inbuilt phase module, \texttt{datafile}, having the name of the binary HDF5 file to store output results and the user supplied permeability and porosity field values, \texttt{kf} and \texttt{phif}, if the inbuilt random field module is not used. Internally it passes the \texttt{filename} to the initialization class as \texttt{IP.ModelInitialization(filename)} creating the attribute repository file \texttt{V.py} which contains domain information along with all other necessary physical parameters. This class also initializes a dictionary, \texttt{Simulate.attr} used for storing the values of concentration, density, viscosity and diffusion coefficient computed at each iteration and making it available to the user for access. If the boolean variable \texttt{UsePhaseModule} corresponds to false, the \texttt{Simulate} class checks for the end-member values of viscosity and density from \texttt{V.py} and based on local concentration, \texttt{c}, declares interpolation functions, \texttt{Simulate.mu\textunderscore func(c)} and \texttt{Simulate.rho\textunderscore func(c)}, to get local cell-wise values. Conversely, if \texttt{UsePhaseModule} is true, the \texttt{Simulate.attr} dictionary is initialized with attributes computed from the inbuilt phase module. The \texttt{Simulate} class also initializes the random fields either supplied by the user as its arguments, or generated from the inbuilt field class, \texttt{Field}, as described in the previous section. It also links to another class \texttt{SpeciesInfo} which contains information about the solubility of $\mathrm{CO_2}$ and mole fraction of different species in the brine. To start the simulation, the user should call the class function \texttt{ParticleTracker(steps,realizations,intervals,params)} in \texttt{Simulate}, where the arguments respectively are number of simulation steps, spatio-temporal evolution of plume dynamics for a single $k-\phi$ field, alternate periods where the user wants to store the data in years and user-specified thermodynamic parameters belonging to \texttt{Simulate.attr} list which is computed form phase equilibrium models other than the one used here. The initial condition is configured by calling the \texttt{\textunderscore configure\textunderscore init\textunderscore condition()} function at the start of every simulation each time the \texttt{ParticleTracker} function is invoked.  The initial condition computes the flow field for the Lagrangian particle simulation to work. Solving the flow field is handled by the \texttt{FlowSolver(field)} class which takes in the random permeability field as the only argument and contains the attributes --- pressure, vorticity and horizontal and vertical flux components. Simulations pertaining to constant viscosity coefficients and variable viscosity coefficients are handled separately. Functions \texttt{\textunderscore AssemblePressureCoefficientMatrix\textunderscore cv()} and \texttt{\textunderscore AssembleCoefficientMatrix\textunderscore rhs(r, PL, PR)} are designed to handle cases arising from constant $\mathrm{CO_2}$--brine viscosity. The former assembles the left hand side of the pressure coefficient matrix and the later assembles the right hand side depending on the density, \texttt{r} and surface pressure values at the boundary, \texttt{PL} and \texttt{PR}. The left hand side of the pressure coefficient matrix, for a constant permeability and viscosity field, is invariant and assembled only once and inverted at the start of the simulation. This inversion is carried out by a LU-decomposition technique as discussed earlier. For a variable viscosity case, the pressure coefficient matrix can no longer be precomputed and hence has to be iteratively solved at each step. This is carried out by the function \texttt{\textunderscore GlobalCoefficientMatrix\textunderscore vv()}. The \texttt{solve(r, mu)} method is invoked by passing in arguments density, \texttt{r} and viscosity, \texttt{mu}, to solve for the pressure field and Darcy flux. At every step the particle tracker defines a particle reservoir, uses the flow field to advect the particle cloud according to the random walk method described before, computes the local concentration by binning and re-computes the flow field for the successive iterations. The \texttt{RWPT(c)} class is responsible for solving the transport equation, where the argument \texttt{c} represents the concentration field. Advection of the particles requires the knowledge of local dispersion tensor, which is computed inside \texttt{CellwiseDispersion()} function present inside \texttt{FlowSolver(field)} class. To disperse particle cloud as per equation \ref{eq16}, the \texttt{disperse(Dxx, Dxy, Dyx, Dyy, phi, vx, vy, plst)} function is invoked which solves the SDE by taking into account the cell-wise disperson tensor (\texttt{Dxx}, \texttt{Dxy}, \texttt{Dyx}, \texttt{Dyy}), local Darcy flux (\texttt{vx}, \texttt{vy}) and porosity (\texttt{phi}) values interpolated at particle locations (\texttt{plst}).    

At each iteration or specified intervals, the simulation results are exported in binary format and stored in the memory. As indicated in figure \ref{fig:10}, these datasets are divided into two categories -- temporally variable, consisting mainly of computed scalar and vector fields like concentration, density, pressure, velocity, vorticity and particle plume configuration that changes at each simulation step and temporally invariable attributes like the domain information consisting of Eulerian nodes and the random $k-\phi$ field that are initialized only once at the start of the simulation.

\section{Summary and Conclusion}  %% \conclusions[modified heading if necessary]
To our knowledge this is the first particle tracking flow solver solely dedicated for usage in the context of deep carbon sequestration and storage. It incorporates the recent thermodynamics models based on $\mathrm{CO_2}$-$\mathrm{H_2O}$-$\mathrm{NaCl}$ systems to compute different phase parameters and couples it in a computationally efficient way with a flow and transport solver. The overall module is customizable and can be used my experimentalists and modellers alike for benchmarking and can also be tailored to solve specific problems. It can efficiently perform large scale simulations of subsurface flow and transport of $\mathrm{CO_2}$--brine mixture at reservoir conditions.

%% The following commands are for the statements about the availability of data sets and/or software code corresponding to the manuscript.
%% It is strongly recommended to make use of these sections in case data sets and/or software code have been part of your research the article is based on.

\codeavailability %% use this section when having only software code available

PyDDC numerical software is a part of an open-source project developed solely in Python. The code can be either downloaded from \url{https://github.com/TectoArc/PyDDC.git} or from \url{https://doi.org/10.5281/zenodo.11114009} and its dependencies can be directly found in the README.md file. It requires a Python version of at least 3.11.0 and can run on any platform for which its package dependencies are available, including Windows, Linux and macOS. 

% \dataavailability{TEXT} %% use this section when having only data sets available

% \codedataavailability{TEXT} %% use this section when having data sets and software code available

% \sampleavailability{TEXT} %% use this section when having geoscientific samples available

% \videosupplement{TEXT} %% use this section when having video supplements available

% \appendix
% \section{}    %% Appendix A

% \subsection{}     %% Appendix A1, A2, etc.

% \noappendix       %% use this to mark the end of the appendix section. Otherwise the figures might be numbered incorrectly (e.g. 10 instead of 1).

% %% Regarding figures and tables in appendices, the following two options are possible depending on your general handling of figures and tables in the manuscript environment:

% %% Option 1: If you sorted all figures and tables into the sections of the text, please also sort the appendix figures and appendix tables into the respective appendix sections.
% %% They will be correctly named automatically.

% %% Option 2: If you put all figures after the reference list, please insert appendix tables and figures after the normal tables and figures.
% %% To rename them correctly to A1, A2, etc., please add the following commands in front of them:

% \appendixfigures  %% needs to be added in front of appendix figures

% \appendixtables   %% needs to be added in front of appendix tables

% %% Please add \clearpage between each table and/or figure. Further guidelines on figures and tables can be found below.

\authorcontribution{The first author, Sayan Sen, took the lead role in this project by developing the numerical software, collecting resources and writing the original draft. The corresponding author, Dr. Scott K. Hansen is the project administrator who handled conceptualization, supervision, writing review and also assisted with the methodology.} %% this section is mandatory

\competinginterests{The authors declare that they have no conflict of interest.} %% this section is mandatory even if you declare that no competing interests are present

% \disclaimer{TEXT} %% optional section

\begin{acknowledgements}
SKH holds the Helen Unger Career Development Chair in Desert Hydrogeology. This project is partially supported by the Israel Science Foundation (ISF), grant number - 1872/19. 
\end{acknowledgements}

%% REFERENCES

%% The reference list is compiled as follows:
% 0\setbibliographystyle{apa}
% \bibliography{ref.bib}

% \begin{thebibliography}{ref.bib}

% \bibitem[Hansen et al. (2018)]{\{Direct breakthrough curve prediction from statistics of heterogeneous conductivity fields}
% REFERENCE 1

% \bibitem[AUTHOR(YEAR)]{LABEL2}
% REFERENCE 2

% \end{thebibliography}

%% Since the Copernicus LaTeX package includes the BibTeX style file copernicus.bst,
%% authors experienced with BibTeX only have to include the following two lines:
%%
\bibliographystyle{copernicus}
\bibliography{template.bib}

\end{document}